\documentclass[fleqn,usenatbib]{mnras}
\usepackage{fix-cm}

\usepackage{newtxtext,newtxmath}
\usepackage{CJKutf8}
\usepackage{xhfill}% http://ctan.org/pkg/xhfill
\newcommand{\ditto}[1][.4pt]{\xrfill{#1}~\textquotedbl~\xrfill{#1}}
\usepackage[normalem]{ulem} 
\usepackage[T1]{fontenc}
\defcitealias{prime1}{Primeval~I}
\defcitealias{prime2}{Primeval~II}
\defcitealias{prime3}{Primeval~III}
\defcitealias{prime4}{Primeval~IV}
\defcitealias{prime5}{Primeval~V}
\defcitealias{prime6}{Primeval~VI}
\defcitealias{prime7}{Primeval~VII}
\DeclareRobustCommand{\VAN}[3]{#2}
\let\VANthebibliography\thebibliography
\def\thebibliography{\DeclareRobustCommand{\VAN}[3]{##3}\VANthebibliography}

\usepackage{graphicx}	% Including figure files
\usepackage{amsmath}	% Advanced maths commands

\title[Benchmark brown dwarfs -- I]{Benchmark brown dwarfs -- I. A blue M2 + T5 wide binary and a probable young [M4 + M4] + [T7 + T8] hierarchical quadruple}
\author[Z. H. Zhang et al.]{Z. H. Zhang (\begin{CJK*}{UTF8}{gbsn}张曾华\end{CJK*}),$^{1,2}$\thanks{E-mail:
zz@nju.edu.cn} F. Navarete,$^{3}$ M. C. G\'alvez-Ortiz,$^{4}$ H. R. A. Jones,$^{5}$ A. J. Burgasser,$^{6}$ P. Cruz,$^{4}$    
\newauthor
F. Marocco,$^{7}$ N. Lodieu,$^{8,9}$ Y. Shan,$^{10}$ B. Gauza,$^{11}$ R. Raddi,$^{12}$ M. R. Huang (\begin{CJK*}{UTF8}{gbsn}黄缪锐\end{CJK*}),$^{1,2}$  R. L. Smart,$^{13}$  
\newauthor
S. Baig,$^{5}$ G. Cheng$^{5}$ and D. J. Pinfield$^{5}$    \\
$^{1}$School of Astronomy and Space Science, Nanjing University, 163 Xianlin Avenue, Nanjing 210023, China \\
$^{2}$Key Laboratory of Modern Astronomy and Astrophysics, Nanjing University, Ministry of Education, Nanjing 210023, China \\
$^{3}$Laborat\'orio Nacional de Astrof\'isica (LNA/MCTI), Rua dos Estados Unidos, 154, 37504-364, Itajubá, MG, Brazil \\
$^{4}$Centro de Astrobiolog{\'i}a (CAB), CSIC-INTA, Camino Bajo del Castillo s/n, E-28692, Villanueva de la Ca{\~n}ada, Madrid, Spain \\
$^{5}$Centre for Astrophysics Research, University of Hertfordshire, Hatfield, Hertfordshire AL10 9AB, UK \\
$^{6}$Department of Astronomy \& Astrophysics, University of California San Diego, La Jolla, CA 92093, USA \\
$^{7}$IPAC, Caltech, Mail Code 100-22, Caltech, 1200 E. California Boulevard, Pasadena, CA 91125, USA \\
$^{8}$Instituto de Astrof{\'i}sica de Canarias, E-38205 La Laguna, Tenerife, Spain \\
$^{9}$Universidad de La Laguna, Dept. Astrof{\'i}sica, E-38206 La Laguna, Tenerife, Spain \\
$^{10}$Centre for Planetary Habitability, Department of Geosciences, University of Oslo, Sem Saelands vei 2b, 0315 Oslo, Norway \\
$^{11}$Janusz Gil Institute of Astronomy, University of Zielona G\'ora, Lubuska 2, PL-65-265 Zielona G\'ora, Poland \\
$^{12}$Universitat Polit\`ecnica de Catalunya, Departament de F\'isica, c/ Esteve Terrades 5, E-08860 Castelldefels, Spain \\
$^{13}$Istituto Nazionale di Astrofisica, Osservatorio Astronomico di Torino, Strada Osservatrio 20, I-10025 Pino Torinese, Italy \\
}

\date{Accepted 2025 May 30. Received 2025 May 30; in original form 2025 January 28}

\pubyear{\the\year{}}

\begin{document}
\label{firstpage}
\pagerange{\pageref{firstpage}--\pageref{lastpage}}
\maketitle

% Abstract of the paper
\begin{abstract}
Benchmark brown dwarfs in wide binary systems are crucial for characterizing substellar objects and calibrating atmospheric and evolutionary models. However, brown dwarf benchmarks with subsolar metallicity, very cool temperatures, or suitability for dynamical mass measurements are rare, limiting our understanding across the full range of mass, age, and metallicity. We present the discovery of two new multiple systems containing T dwarf companions, identified through a targeted search using CatWISE2020 and {\sl Gaia} catalogues. L 122-88 AB is a wide binary comprising a mildly metal-poor M2 dwarf and a T5 dwarf, separated by 215.6 arcsec at a distance of 33.106$\pm$0.014 pc. Atmospheric model fitting to the near infrared spectrum of L 122-88 A suggests a mildly metal-poor composition ([Fe/H] = $-$0.2). UPM J1040$-$3551 AB is a candidate hierarchical quadruple system at 25.283$\pm$0.013 pc, consisting of a likely astrometric binary of two M4 dwarfs and a probable unresolved spectral binary of T7 and T8 dwarfs, separated by 65.48 arcsec from the primary. The H$\alpha$ emission detected in UPM J1040$-$3551 A indicates an age range of 0.3-2.0 Gyr. This age estimate suggests that the T8 component has a mass between 9 and 28 Jupiter masses, potentially classifying it as a planetary-mass object. These systems augment the sample of benchmark brown dwarfs, particularly in the underexplored regime of cool temperature, providing valuable opportunities for refining our understanding of substellar objects.

%This is a simple template for authors to write new MNRAS papers.
%The abstract should briefly describe the aims, methods, and main results of the paper.
%It should be a single paragraph not more than 250 words (200 words for Letters).
%No references should appear in the abstract.

\end{abstract}

% Select between one and six entries from the list of approved keywords.
% Don't make up new ones.


%%%%%%%%%%%%%%%%%%%%%%%%%%%%%%%%%%%%%%%%%%%%%%%%%%

%%%%%%%%%%%%%%%%% BODY OF PAPER %%%%%%%%%%%%%%%%%%

\section{Introduction}
Brown dwarfs (BDs) are the most recently discovered major population on the Hertzsprung-Russell diagram. They mostly form like stars but lack sufficient mass to sustain steady hydrogen fusion. Consequently, BDs represent the low-mass tail of the initial mass function and are crucial for constraining it \citep{kirk24}. Bridging the gap between stars and planets, BDs share similar physical and atmospheric properties with gas giant planets \citep{hatz15}, providing essential references for the characterization of exoplanets.

To address the scientific objectives outlined above, samples of well-characterized BDs are required. However, BDs are challenging to characterize due to their faintness, cool temperatures, and the mass–age degeneracy inherent in their evolution. Without steady hydrogen burning, BDs cool over time, altering their spectral features and types \citep[e.g. fig. 8 of][]{burr01}. Therefore, benchmark BD samples with known mass, age, metallicity, and luminosity, spanning a wide parameter space, are needed to test atmospheric and evolutionary models. This need is especially pressing in the era of ongoing and upcoming deep surveys such as {\sl Euclid} \citep{laur11}, the {\sl Roman Space Telescope} \citep[][]{sper15}, the Rubin Observatory \citep[][]{lsst17}, and the {\sl Chinese Space Station Telescope} \citep[{\sl CSST};][]{zhan11}. 

BDs in nearby open clusters or moving groups with well-established ages serve as valuable age benchmarks [e.g., Teide 1, \citep{rebo95} and Luhman 16 AB, \citep{luhm13}]. However, these clusters and groups are predominantly young ($\lesssim$ 1 Gyr) and relatively distant ($\gtrsim$ 100 pc). In contrast, field BDs constitute the majority of the BD population, and span a broader range of ages and metallicities, with many residing in the solar neighbourhood. Yet, direct measurements of age or mass for individual field BDs remain elusive. None the less, BDs that are wide companions to well-characterized stars provide ideal benchmarks [e.g., Gliese 229 B, \citep{naka95} and $\varepsilon$ Indi B, \citep{scho03,mcca04,king10}]. Binary systems are expected to have coeval formation, sharing common age, metallicity, and distance -- parameters more readily determined for stars than for BDs. This characteristic makes such systems particularly valuable for BD studies.

The T dwarf spectral class was established to classify BDs cool enough to exhibit methane absorption in their near-infrared (NIR) spectra \citep{burg02,geba02}. T dwarfs have effective temperatures ($T_{\rm eff}$) between $\sim$500 and 1300 K \citep{kirk21}. They constitute the majority of field BDs \citep[e.g.][]{prime6}. However, T dwarfs are relatively rare in observations due to their cool temperatures and low luminosities. About 900 spectroscopically confirmed T dwarfs are known to date, mostly within 30 pc of the Sun  \citep[see summary in][]{best24}. Two-thirds of known T dwarfs have spectral types of T5–T9, owing to the rapid evolution of BDs through the L/T transition \citep[e.g.][]{best21}.

Currently, there are about 30 T dwarfs known with stellar companions \citep[summarized in][]{best24}, with significant contributions from the {\sl Wide-field Infrared Survey Explorer} \citep[{\sl WISE};][]{wrig10}. To expand the parameter space of benchmark T dwarfs and improve their characterization, we searched for late-type T dwarfs with wide stellar companions using the CatWISE2020 catalogue \citep{maro21} and the third data release (DR3) of {\sl Gaia} \citep{gaiadr3}. In this paper, we present the discovery of a wide binary of mildly metal-poor M2 + T5 dwarfs and a probable young hierarchical quadruple system of [M4 + M4] + [T7 + T8] dwarfs. Section \ref{ssel} describes the selection of wide binary candidates. Spectroscopic observation and characterization of these two systems are presented in Sections \ref{sspec} and \ref{scha}, respectively. A summary and conclusions are presented in Section \ref{scon}.

\section{Selection of wide binaries}
\label{ssel}

Wide binaries ($\gtrsim$ 100 au) exhibit negligible orbital motion over decadal time-scales due to their extremely long orbital periods ($\gtrsim$ 1 kyr). In the solar neighbourhood ($\lesssim$ 100 pc), the proper motions (PMs) of wide binary components significantly exceed their orbital motions, resulting in apparent common PM over decades. This characteristic makes common PM an effective method for identifying nearby wide binaries \citep[e.g.,][]{prime7,prime8}.

T dwarfs emitting primarily in the infrared, only those in the solar neighbourhood are detectable by large-scale infrared surveys. The CatWISE catalogue, providing mid-infrared photometry and PMs for nearby objects, serves as a powerful tool for identifying nearby T dwarf wide binaries. 

We focused on late-type T dwarfs, which exhibit redder $W1-W2$ colour (equation \ref{eq:color}) and suffer less contamination, e.g. from poorly measured background objects, than earlier types. Candidate selection from CatWISE employed the following criteria:
\begin{eqnarray}
    \label{eq:color}
    W1 - W2 > 1.8 \\
    \label{eq:snr}
    W1\_snr, W2\_snr > 5 \\
    \label{eq:pm}
    PM > 100~{\rm mas~yr^{-1}} \\
    \label{eq:pmerr}
    PM_{RA}\_error, PM_{Dec}\_error < 50~{\rm mas~yr^{-1}} \\
    \label{eq:b}
    |b| > 10~{\rm deg} \\
    \label{eq:nb}
    nb = 1 \\
    \label{eq:ab}
    ab\_flags = 00  
\end{eqnarray}
In our selection, we required signal-to-noise ratio (SNR) of better than 5.0 for $W1$ and $W2$ detections (equation \ref{eq:snr}). SNR of 5.0 corresponds to an error of $\sim$0.22 in W1 or W2 magnitudes. 
To mitigate contamination from background stars in the low-resolution (2.75 arcsec pixel$^{-1}$) unblurred coadds of the {\sl WISE} images \citep[unWISE;][]{lang14} used for CatWISE, we excluded the Galactic plane (equation \ref{eq:b}). We select objects without identified artefacts in both $W1$ and $W2$ band unWISE images (equation \ref{eq:ab}) and with only one blend component used in the fit (equation \ref{eq:nb}). Approximately 9,540 objects in CatWISE met these criteria (Eqs \ref{eq:color}--\ref{eq:ab}).

We then searched for wide companions of these objects in the {\sl Gaia} Catalogue of Nearby Stars \citep[GCNS;][]{smar21} using common PM. Our common PM pair criteria were:
\begin{eqnarray}
    \label{eq:cpm}
    |PM_{RA1} - PM_{RA2}|, |PM_{Dec1} - PM_{Dec2}| < 40~{\rm mas~yr^{-1}} \\
    \label{eq:dis}    
    distance < 40~pc 
\end{eqnarray}
Given the relatively large PM errors in CatWISE, we required significant PM (equation \ref{eq:pm}) but relaxed the PM constraint in the GCNS cross-match (equation \ref{eq:cpm}). The distance limit of 40 pc (equation \ref{eq:dis}) corresponds to the $W1$-band detection limit of CatWISE for late T dwarfs \citep[e.g. table 6 of][]{prime6}. We limited the binary separation to within 10 arcmin~(corresponding to 24,000 au at 40 pc), beyond which the number of random contaminations arises, partially due to the large PM errors in CatWISE.  

\begin{figure}
	\includegraphics[width=\columnwidth]{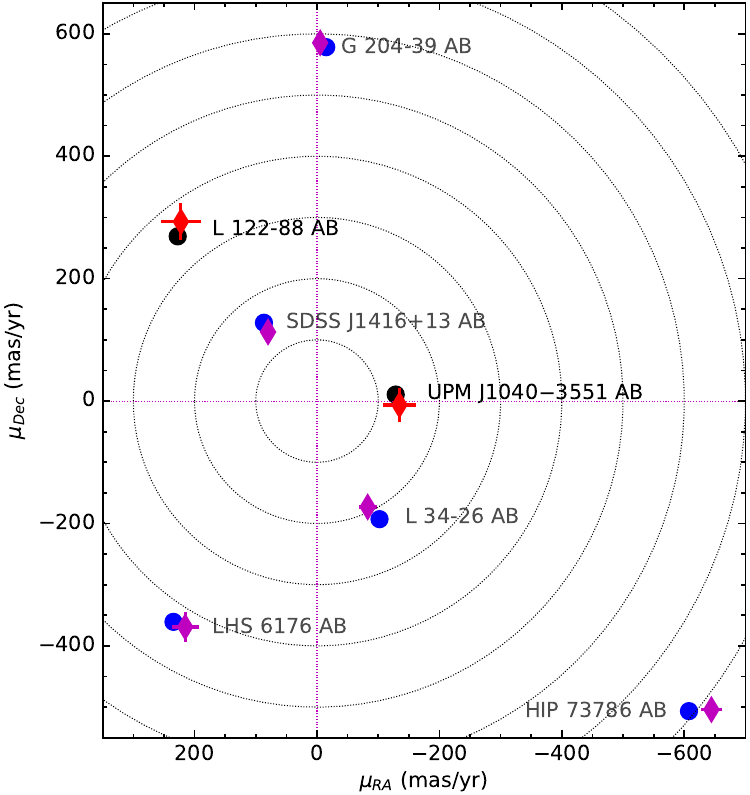}
    \caption{PMs of the newly discovered systems L 122-88 AB and UPM J1040$-$3551 AB, alongside the previously known systems: G 204-39 AB (M3+T6.5), SDSS J1416+13 AB (sdL7+sdT7.5), LHS 6176 AB (K8V+T8p), HIP 73786 AB (sdM4+sdT5.5), and L 34-26 + COCONUTS-2b (M3V+T9) that were recovered in our search. Gliese 570 A \& D (K4V + T7.5) are not shown because their PMs (2 arcsec~yr$^{-1}$) are far exceeding the axis ranges. Black/blue dots represent {\sl Gaia} PMs of primary stars (error bars are smaller than the symbol size), while red/magenta diamonds indicate CatWISE PMs of T dwarf secondaries.}
    \label{fpm}
\end{figure}

Our search yielded 15 candidate common PM pairs. After visual inspection using data from the Dark Energy Survey \citep{desdr2} and the Dark Energy Spectroscopic Instrument (DESI) Legacy Imaging Surveys \citep{dey19}, we excluded seven pairs of which the faint component has poor unWISE images, has Dark Energy Camera (DECam) deep $z$-band non-detection ($z \gtrsim 22.5$), or has no DECam observation and non-detection by other shallower surveys. The remaining eight T dwarf candidates, all with clear DECam $z$-band detections, comprise two newly identified wide binaries and six known wide systems [Gliese 570 A \& D, K4V + T7.5, \citep{burg00}; G 204-39 AB, M3+T6.5, \citep{fahe10}; SDSS J1416+13 AB, sdL7+sdT7.5, \citep{scho10,burn10}; HIP 73786 AB, sdM4+sdT5.5, \citep{murr11,prime6}; LHS 6176 AB, K8V+T8p, \citep{burn13}; L 34-26 + COCONUTS-2b, M3V+T9, \citep{zhan21}]. Fig. \ref{fpm} shows the PMs of these common PM pairs. 

\begin{figure*}
\includegraphics[width=\textwidth]{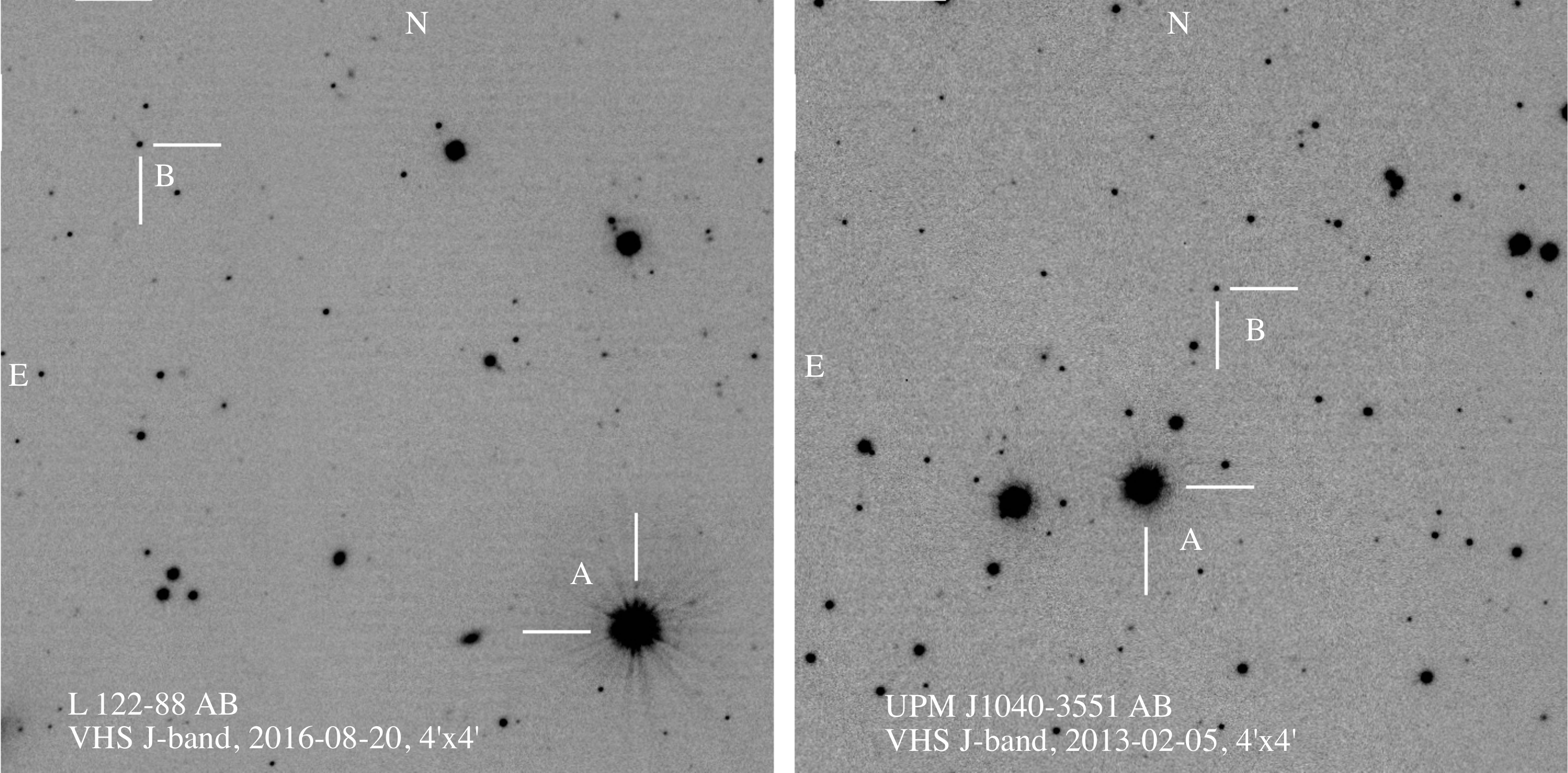}
    \caption{VISTA $J$-band images of the fields surrounding L 122-88 AB (left panel) and UPM J1040$-$3551 AB (right panel), which are indicated with white bars of 20 arcsec in length. Both images cover a 4 arcmin field of view, oriented with north up and east to the left. Observation dates are shown in yyyy-mm-dd format. L 122-88 AB is moving towards the north-east with a PM of 0.368 arcsec~yr$^{-1}$, while UPM J1040$-$3551 AB is moving westwards with a PM of 0.135 arcsec~yr$^{-1}$ (see Fig. \ref{fpm} for PM details). }
    \label{ffc}
\end{figure*}

The two new binaries, L 122-88 AB and UPM J1040$-$3551 AB have been independently reported recently. \citet{maro24} classified UCAC3 52-1038 B (i.e. L 122-88 B) as a T6 dwarf based on photometric colours, and classified  UCAC3 52-1038 A (i.e. L 122-88 A) as an M2 dwarf based on its low-resolution optical spectrum observed on 2021 December 25. \citet{roth24} classified UPM J1040–3551 A as an M3.8 dwarf based on photometric colours, and classified UPM J1040–3551 B as T7 dwarf based on a low-resolution NIR spectrum observed on 2023 June 30. Figure \ref{ffc} shows the finding charts of these two new binaries. Their properties are listed in Table \ref{ttb}. The binary nature of these two newly identified pairs is corroborated by the concordance between the {\sl Gaia} parallax-derived distances of the stellar components and the spectroscopically determined distances of their BD companions (Section \ref{sstd}). 

L 122-88 AB and UPM J1040$-$3551 AB are located at distances of 33 and 25 pc respectively. They were missed by earlier searches with {\sl WISE} and AllWISE \citep{wrig10,main11}, possibly because T dwarf candidates within 20 pc were prioritized in those spectroscopic follow up programmes, to achieve a more complete volume-limited sample \citep[e.g.,][]{kirk11,kirk21,mace13}. 
Another seven known T dwarf binaries were selected as T dwarf candidates by equations (\ref{eq:color}--\ref{eq:ab}), but were not recovered as wide binaries mainly because of large uncertainties of their CatWISE PMs. We recovered most of them by double the limits in equations (\ref{eq:pm}) and (\ref{eq:pmerr}), but no new binary T dwarf candidate was found.

\begin{table*}
 \centering
  \caption[]{Properties of L 122-88 AB and UPM J1040$-$3551 AB.}
\label{ttb}
  \begin{tabular}{l c c c}
\hline
Primary & L 122-88 A & UPM J1040$-$3551 Aa/Ab & Ref. \\
CWISE &  J003124.30$-$641354.9   &  J104055.33$-$355130.9   \\
\hline 
{\sl Gaia} DR3 & 4708465482876995712   & 5443160355149610368 & (1)  \\
 $\alpha$ (2016)   & $00^{\rm h}31^{\rm m}24\fs32$ & $10^{\rm h}40^{\rm m}55\fs32$ & (1) \\
 $\delta$ (2016)   &  $-64\degr13\arcmin54\farcs8$ & $-35^{\rm h}51^{\rm m}30\fs9$   & (1)   \\
Spectral type & M2 & M4 + M4 & (2) \\
Age (Gyr) & 2.5-9.9 & 0.3-2.0 & (3,2)\\
RUWE & 1.32 & 1.47 & (1) \\
$G$ & 12.209   & 13.230  & (1) \\
$G_{\rm BP}$ & 13.346   & 14.863  & (1) \\
$G_{\rm RP}$ & 11.151   & 11.997  & (1) \\
$J$ (2MASS) & 9.812 $\pm$ 0.026 & 10.302 $\pm$ 0.023 & (4) \\
$H$ (2MASS) & 9.248 $\pm$ 0.023 & 9.676 $\pm$ 0.022  & (4)\\
$K$ (2MASS) & 8.991 $\pm$ 0.019 & 9.446 $\pm$ 0.023 & (4) \\
$W1$ & 8.868 $\pm$ 0.012 & 9.280 $\pm$ 0.012 & (5)\\
$W2$ & 8.729 $\pm$ 0.007 & 9.087 $\pm$ 0.007 & (5) \\
$\varpi$ (mas) & 30.206 $\pm$ 0.013 & 39.552 $\pm$ 0.020 & (1)  \\
Distance (pc) & 33.106 $\pm$ 0.014 & 25.283 $\pm$ 0.013  & (1) \\
$\mu_{\rm RA}$ (mas yr$^{-1}$) & 227.556 $\pm$ 0.014 & $-$128.580 $\pm$ 0.019 & (1) \\
$\mu_{\rm Dec}$ (mas yr$^{-1}$) & 269.121 $\pm$ 0.014 & 10.646 $\pm$ 0.018 & (1) \\
$V_{\rm tan}$ (km s$^{-1}$) & $55.304\pm0.023$ & $15.462\pm0.008$  & (1) \\
RV (km s$^{-1}$) & 15.63 $\pm$ 0.31 & $-$0.09 $\pm$ 3.72  & (1)\\
$M_K$ & 6.391 $\pm$ 0.020 & 7.432 $\pm$ 0.024 & (2) \\
$T_{\rm eff}$ (K) & 3618 $\pm$ 157 & 3244 $\pm$ 157  & (6) \\ 
$T_{\rm eff}$ (K) & 3500 $\pm$ 50 &  3200 $\pm$ 50 & (2) \\ 
${\rm [Fe/H]}$ & $-0.2\pm0.1$ & $0.0\pm0.1$ & (2) \\ 
Mass (M$_{\odot}$) & 0.378 $\pm$ 0.020 & 0.174 $\pm$ 0.022/0.165 $\pm$ 0.020 & (6,2) \\
\hline 
Companion B & L 122-88 B & UPM J1040$-$3551 Ba/Bb  \\
CWISE &  J003147.79$-$641123.1   &  J104053.42$-$355029.7   \\
\hline 
 $\alpha$ ({\sl WISE})   & $00^{\rm h}31^{\rm m}47\fs79$ & $10^{\rm h}40^{\rm m}53\fs42$  & (3) \\
 $\delta$  ({\sl WISE})   &  $-64\degr11\arcmin23\farcs1$ & $-35^{\rm h}50^{\rm m}29\fs7$  & (3)  \\
 Mean MJD (d) & 57265.87 & 57124.73 & (3) \\
Spectral type & d/sdT5   & T7 + T8  & (2)\\
$r$ (DECaPS) & 25.149 $\pm$ 0.427 &  ---   & (7)\\
$i$ (DECaPS) & 23.824 $\pm$ 0.216 &   --- & (7)\\
$z$ (DECaPS) & 21.107 $\pm$ 0.035 & 21.133 & (7) \\
$Y$ (DECaPS) & 19.821 $\pm$ 0.046 &  --- & (7) \\
$J$ (VHS) & 17.601 $\pm$ 0.015 & 17.393 $\pm$ 0.022 & (8) \\
$Ks$ (VHS) & 18.417 $\pm$ 0.239 & 17.344 $\pm$ 0.128 & (8) \\
$W1$ & 17.380 $\pm$ 0.055 & 16.773 $\pm$ 0.041 & (5) \\
$W2$ & 15.351 $\pm$ 0.029 & 14.940 $\pm$ 0.025 & (5) \\
Distance (pc) &  45$^{+21}_{-17}$  &  $^a$20$^{+9}_{-8}$, $^b$29$^{+11}_{-14}$   & (2) \\
$\mu_{\rm RA}$ (mas yr$^{-1}$) & 222 $\pm$ 33 & $-$135 $\pm$ 27 & (5)\\
$\mu_{\rm Dec}$ (mas yr$^{-1}$) & 293 $\pm$ 30 & $-$6 $\pm$ 28 & (5) \\
Mass (M$_{\odot}$) & 0.05--0.067 & 0.011--0.033/0.009--0.027 & (2) \\
\hline 
Separation (arcsec) & 215.6  & 65.48  & (2) \\
Proj. sep. (au) & 7138.4 $\pm$ 0.5 & 1655.6 $\pm$ 0.3 & (2) \\
$^c$Proj. sep. ($r_{\rm J}$) &  0.034 & 0.008  & (2) \\
$-U$ (J) &  $5.6\times10^{33}$ & $1.3\times10^{34}$ & (2) \\
\hline 
\end{tabular}
\begin{list}{}{}
\item[]References:
(1) \citet{gaiadr3}; (2) This paper; 
(3) \citet{maro24}; (4) \citet{2mass06}; (5) \citet{maro21}; 
(6) \citet{stas19}; 
(7) \citet{dey19}; and (8) \citet{mcma13}.
\item[]$^a$Average spectroscopic distance as a single T7 dwarf based on spectral type -- absolute magnitudes in $J, K, W1, W2$ bands \citep{prime6}.  
\item[]$^b$Average spectroscopic distance as a T7+T8 binary. The extracted apparent magnitude of the T7 companion is used in the calculation. 
\item[]$^c$Projected separation in Jacobi radius ($r_{\rm J}$, see Section \ref{ssbi}).
\end{list}
\end{table*}

\begin{table*}
 \centering
  \caption[]{Summary of the characteristics of the spectroscopic observations made with SOAR and VLT. }
\label{tobs}
  \begin{tabular}{l c c c c c c c l c l}
\hline
    Name  & SpT & UT date & Telescope & Instrument & Slit & Seeing  & Airmass  & Wavelength  & Resolution & $T_{\rm int}$      \\
     & & & & &  (arcsec) &  (arcsec) & & (nm) & ($\lambda/\delta\lambda$) & (s)\\   
\hline
UPM J1040–3551 A & M5 & 2023-12-29 & SOAR & Goodman/400 & 1.0 & 1.45 & 1.04  & 492-897 & 2000 &  $1 \times 75$ \\
UPM J1040–3551 B & T7+T8 & 2024-03-26 & SOAR & TripleSpec & 1.1 & 0.91 & 1.07  & 940-2470 & 3500 & $8 \times 225$  \\
L 122-88 A & M2 & 2023-12-29 & SOAR & Goodman/400 & 1.0 & 1.55 & 1.89  & 492-897 & 2000 & $1 \times 75$ \\
\ditto & \ditto & 2024-06-16 & VLT & X-shooter/UVB & 1.3 & 1.25 & 1.36  & 294-593 & 4000 & $2 \times 64$ \\
\ditto & \ditto & \ditto & \ditto & X-shooter/VIS & \ditto & \ditto & \ditto  & 526-1048 & 6700 & $2 \times 93$ \\
\ditto & \ditto & \ditto & \ditto & X-shooter/NIR & \ditto & \ditto & \ditto  & 983-2480 & 4300 & $4 \times 50$ \\
L 122-88 B & d/sdT5 & 2024-07-11 & SOAR & TripleSpec & 1.1 & 1.24 & 1.35  & 940-2470 & 3500 & $20 \times 200$  \\
\ditto & \ditto & 2024-07-07 & VLT & X-shooter/VIS & 1.2 & 3.58 & 1.32  & 526-1048 & 6700 & $2 \times 1474$ \\
\ditto & \ditto & \ditto & \ditto & X-shooter/NIR & \ditto & \ditto & \ditto  & 983-2480 & 4300 & $10 \times 300$ \\
\ditto & \ditto & 2024-07-13 & VLT & X-shooter/VIS & 1.2 & 1.29 & 1.45  & 526-1048 & 6700 & $2 \times 1474$ \\
\ditto & \ditto & \ditto & \ditto & X-shooter/NIR & \ditto & \ditto &\ditto  & 983-2480 & 4300 & $10 \times 300$ \\
\hline
\end{tabular}
\end{table*}

\section{Spectroscopic observations}
\label{sspec}
We conducted spectroscopic follow-up observations of L 122-88 AB and UPM J1040$-$3551 AB prior to their independent reporting. As spectra for L 122-88 B and UPM J1040$-$3551 A were not available in the literature, we continued our spectroscopic campaign to complete the analysis of these new binary systems. Table \ref{tobs} summarizes the details of our observations.

\subsection{SOAR/Goodman}
\begin{figure*}
	\includegraphics[width=\textwidth]{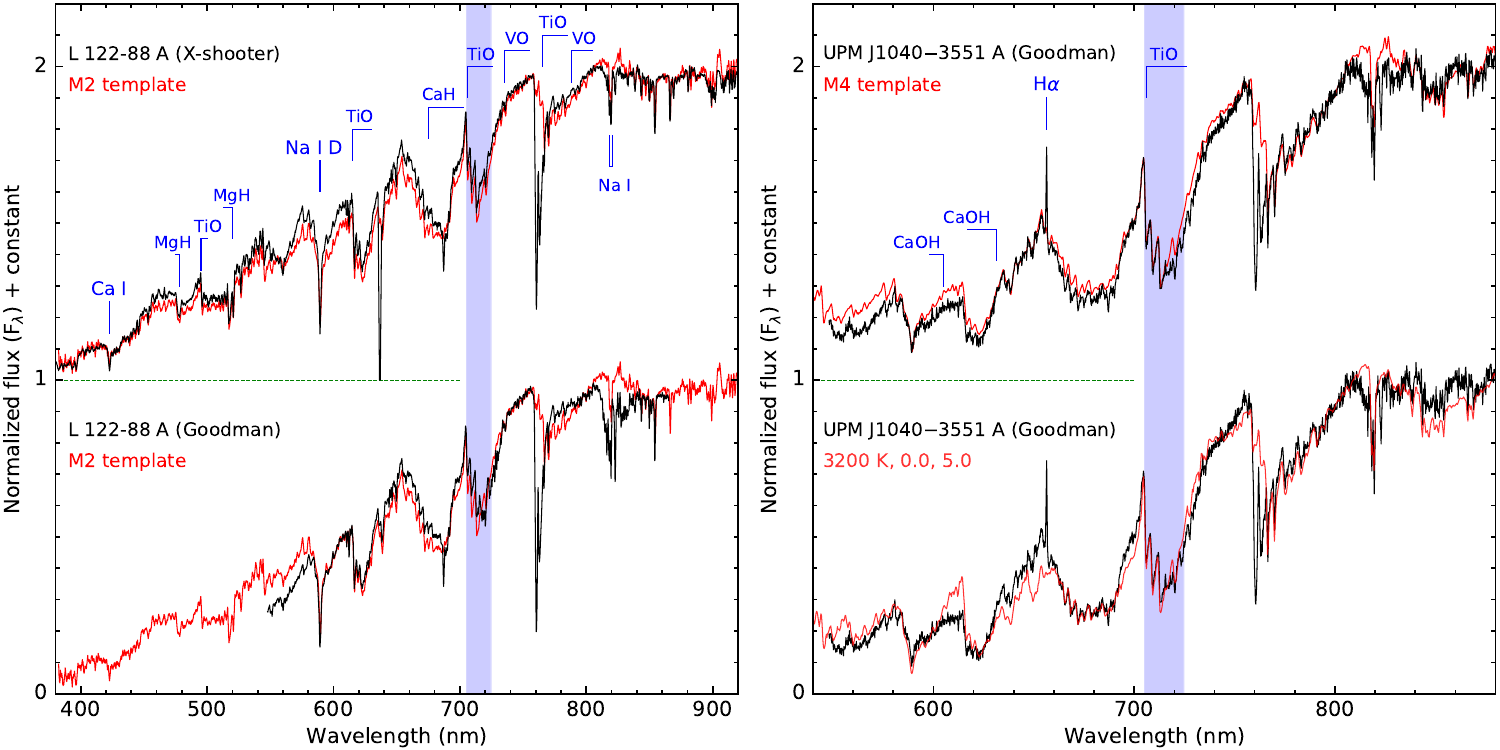}\\
    \caption{Optical spectra of L 122-88 A (left panel) and UPM J1040$-$3551 A (right panel) compared to M2 and M4 templates from \citet{boch07}, respectively. The lower spectrum in the right panel shows the best-fitting BT-Settl model ($T_{\rm eff}$ = 3200 K, [Fe/H] = 0.0, and log $g$ = 5.0) for UPM J1040$-$3551 A, degraded to a resolution of 0.5 nm. Note that telluric absorptions around 760 nm are not corrected for spectra of both primaries.}
    \label{fspeca}
\end{figure*}

L 122-88 A and UPM J1040–3551 A were observed using the Goodman High Throughput Spectrograph \citep[Goodman;][]{clem04} at the Southern Astrophysical Research (SOAR) telescope in long-slit mode on 2023 December 29 (see Fig. \ref{fspeca}). We used a $1\arcsec$ slit and the 400 lines mm$^{-1}$ grism in M2 mode, providing wavelength coverage from 492 to 897 nm with a resolving power of $\sim$2000. Observations were conducted at airmass of 1.89 and seeing of 1.55 for L 122-88 A, and airmass of 1.04 and seeing of 1.45 for UPM J1040–3551 A.

Data reduction was performed using the Goodman High Throughput Spectrograph Pipeline\footnote{\url{https://soardocs.readthedocs.io/projects/goodman-pipeline/}} for basic CCD reduction, including bias subtraction, flat-field correction, and cosmic ray removal. Further spectroscopic processing utilized tasks from the {\scriptsize NOAO} package in {\scriptsize IRAF}, including wavelength calibration, spectral extraction, telluric correction, and flux calibration. HIP 45167 served as the telluric standard star for flux calibration.

\begin{figure*}
\includegraphics[width=0.495\textwidth]{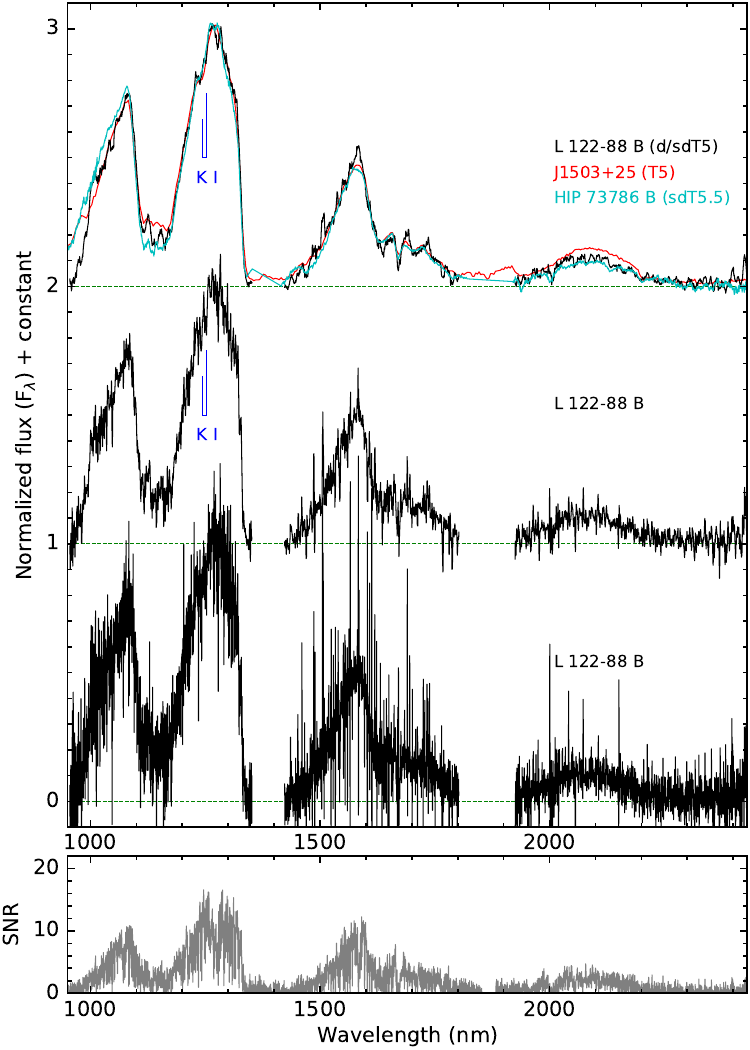} 
\includegraphics[width=0.495\textwidth]{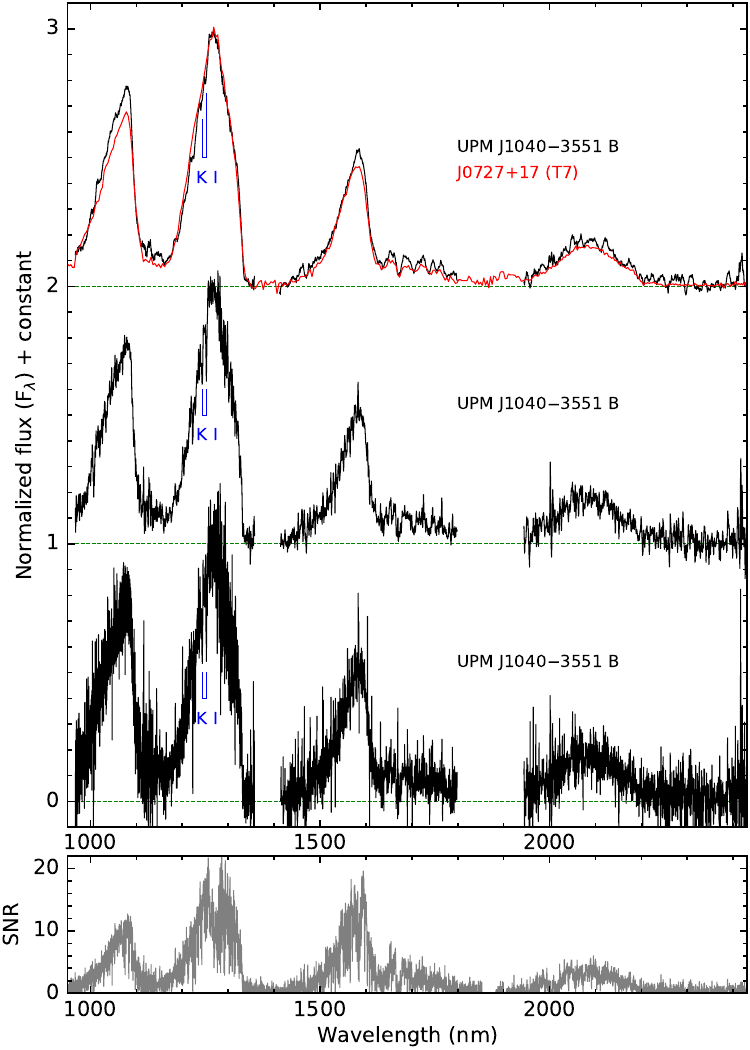}\\
    \caption{NIR spectra of L 122-88 B (left panel) and UPM J1040$-$3551 B (right panel) obtained with TripleSpec. The spectra are shown at original resolution (bottom) and smoothed by 11 pixels (middle) and 51 pixels (top). For comparison, we over plotted the spectra of the sdT5.5 subdwarf HIP 73786B \citep{prime6}, the T5 dwarf standard 2MASS J15031961+2525196 \citep[J1503+25;][]{burg04}, and the T7 dwarf standard 2MASS J0727182+171001 \citep[J0727+17;][]{burg06}.}
    \label{ftspec}
\end{figure*}

\begin{figure*}
	\includegraphics[width=\textwidth]{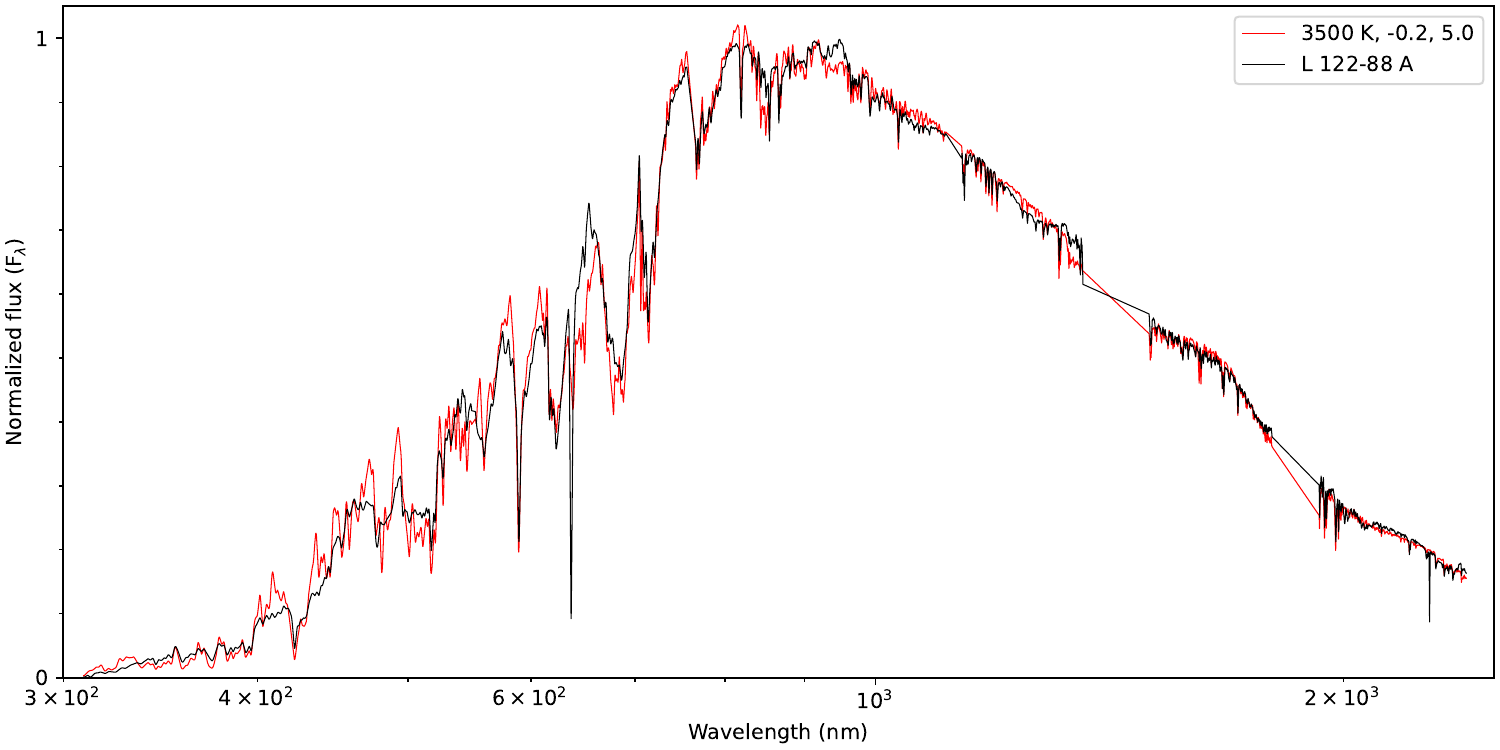}
 \caption{The X-shooter spectrum of L 122-88 A and its best fitting BT-Settl model spectrum ($T_{\rm eff}$ = 3500 K, [Fe/H] = $-0.2$, and log $g$ = 5.0). 
    }
    \label{fj0031axs}
\end{figure*}

\begin{figure*}
\includegraphics[width=\textwidth]{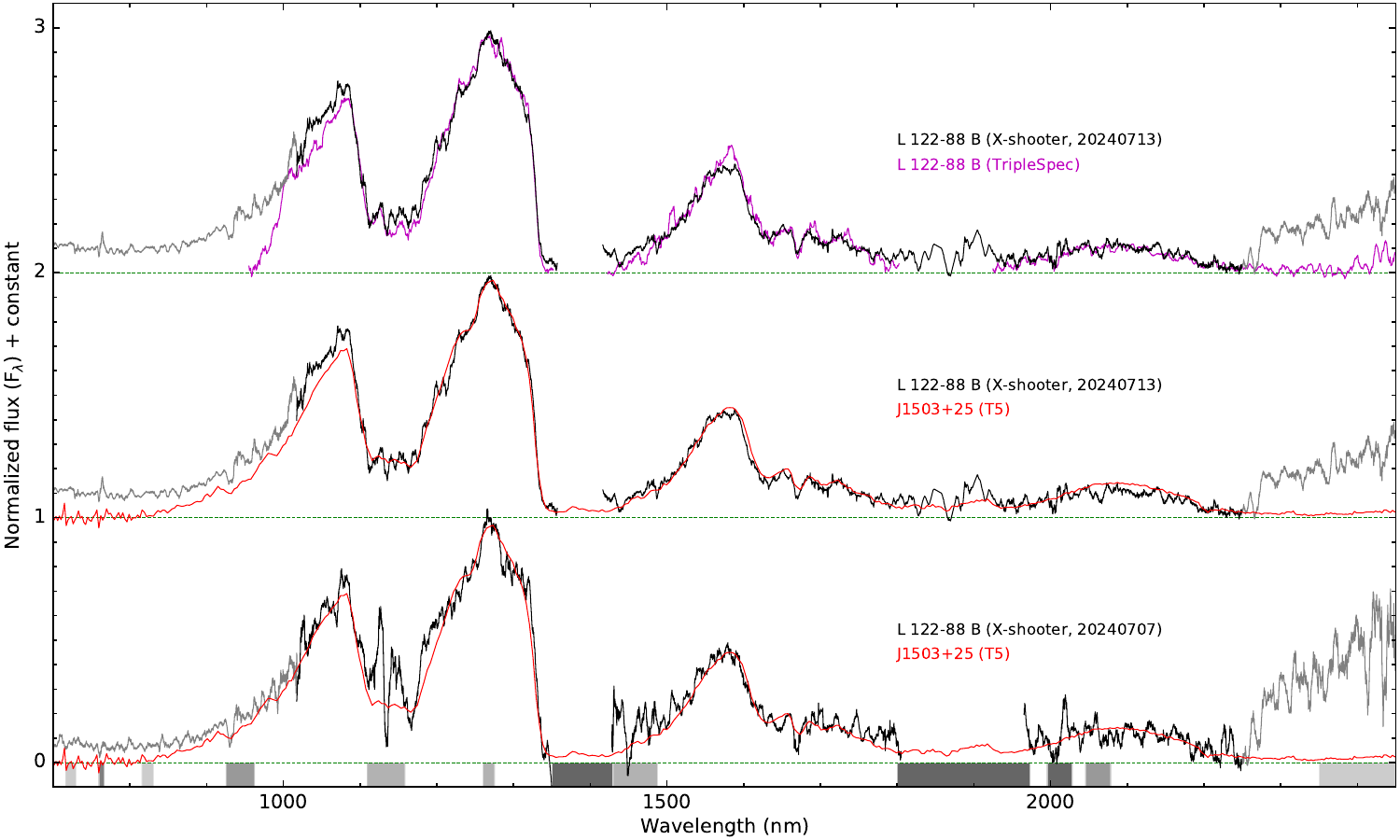} 
    \caption{NIR spectra of L 122-88 B obtained with X-shooter on 2024 July 7 (bottom panel) and July 13 (middle panel).  The spectra have been smoothed by 201 pixels in the VIS arm and 101 pixels in the NIR arm. The July 13 spectrum of L 122-88 B is compared to its TripleSpec spectrum at the top panel. The spectrum of the T5 dwarf standard 2MASS J15031961+2525196 \citep[J1503+25;][]{burg04} is over plotted for comparison. Grey stripes at the bottom indicate telluric regions, which have been corrected in both spectra.}
    \label{fj0031b}
\end{figure*}

\subsection{SOAR/TripleSpec}

UPM J1040$-$3551 B and L 122-88 B were observed with the TripleSpec4.1 NIR Imaging Spectrograph \citep[TripleSpec;][]{schl14} at SOAR. TripleSpec provides simultaneous spectral coverage from 940 to 2470 nm ($R\sim$3500), spanning the entire $YJHK$ photometric range (see Fig. \ref{ftspec}). UPM J1040$-$3551 B was observed on 2024 March 26 (seeing: 0.91 arcsec, airmass: 1.07), while L 122-88 B was observed on 2024 July 11 (seeing: 1.24 arcsec, airmass: 1.35). We adopted the default ABBA dither pattern with eight 225-s exposures for UPM J1040–3551 B and twenty 200-s exposures for L 122-88 B.

Data reduction employed a modified version of the IDL-based Spextool pipeline \citep{Cushing2004}, adapted for SOAR. The process included flat-field correction, wavelength calibration using a CuHeAr arc lamp, removal of emission sky features by subtracting A and B exposures, and one-dimensional spectral extraction. Telluric correction and flux calibration were performed using observations of the standard star HIP 54311. The resulting NIR spectra exhibit peak SNRs of $\sim$10-15 at the centre of the $YJH$ bands.

\subsection{VLT/X-shooter}
The X-shooter spectrograph \citep{vern11} at the Very Large Telescope (VLT) comprises three spectroscopic arms [UV-Blue (UVB), Visible (VIS), and NIR], each optimized for its spectral range (300-2500 nm).
The optical to NIR spectrum of L 122-88 A was obtained using the X-shooter on 2024 June 16 (seeing: 1.5 arcsec, airmass: 1.8). Figure \ref{fj0031axs} presents the combined, smoothed X-shooter spectrum of L 122-88 A.

L 122-88 B was observed with X-shooter on 2024 July 7 (seeing: 3.58 arcsec) and 2024 July 13 (seeing: 1.29 arcsec).
Fig. \ref{fj0031b} shows the X-shooter spectrum of L 122-88 B; noting that the optical spectrum is contaminated by a nearby galaxy, in fact, we note the presence of a non-zero continuum in the 400-800 nm spectral region in which T dwarfs should barely have emitted any flux \citep[e.g.,][]{burg03b}. The nearby galaxy is faint in the $J$ band (see Fig. \ref{ffc}) but brighter than L 122-88 B in the optical. With a PM of 0.368 arcsec yr$^{-1}$ towards the north-east, L 122-88 B has moved to the top of the galaxy in 2024 when the X-shooter spectrum was taken. Meanwhile, the flux of the X-shooter spectrum beyond 2250 nm is poorly extracted/calibrated due to low SNR. These spectra were reduced using the European Southern Observatory Reflex pipeline \citep{freu13}.

\begin{figure}
    \includegraphics[width=\columnwidth]{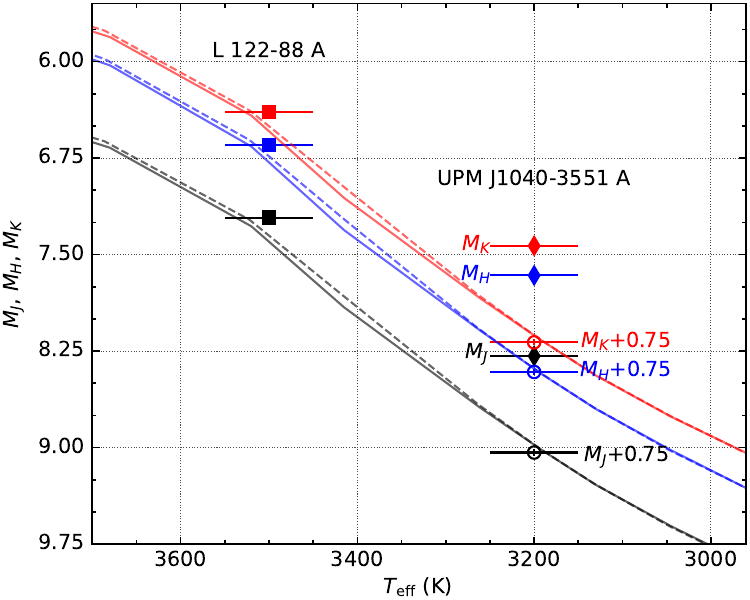}   
    \caption{Observed $M_J$ (black), $M_H$ (blue), and $M_K$ (red) absolute magnitudes of L 122-88 A (filled squires) and UPM J1040$-$3551 A (filled diamonds) compared to 1 Gyr (solid lines) and 8 Gyr (dashed lines) isochrones of solar-metallicity evolutionary models \citep{bara15}. Open circles indicate absolute magnitudes of an individual component of UPM J1040$-$3551 Aab, assumed as an equal-mass binary. Note, errors on absolute magnitudes are much smaller than symbol size.}
    \label{ftjhk}
\end{figure}

\begin{figure*}
\includegraphics[width=\textwidth]{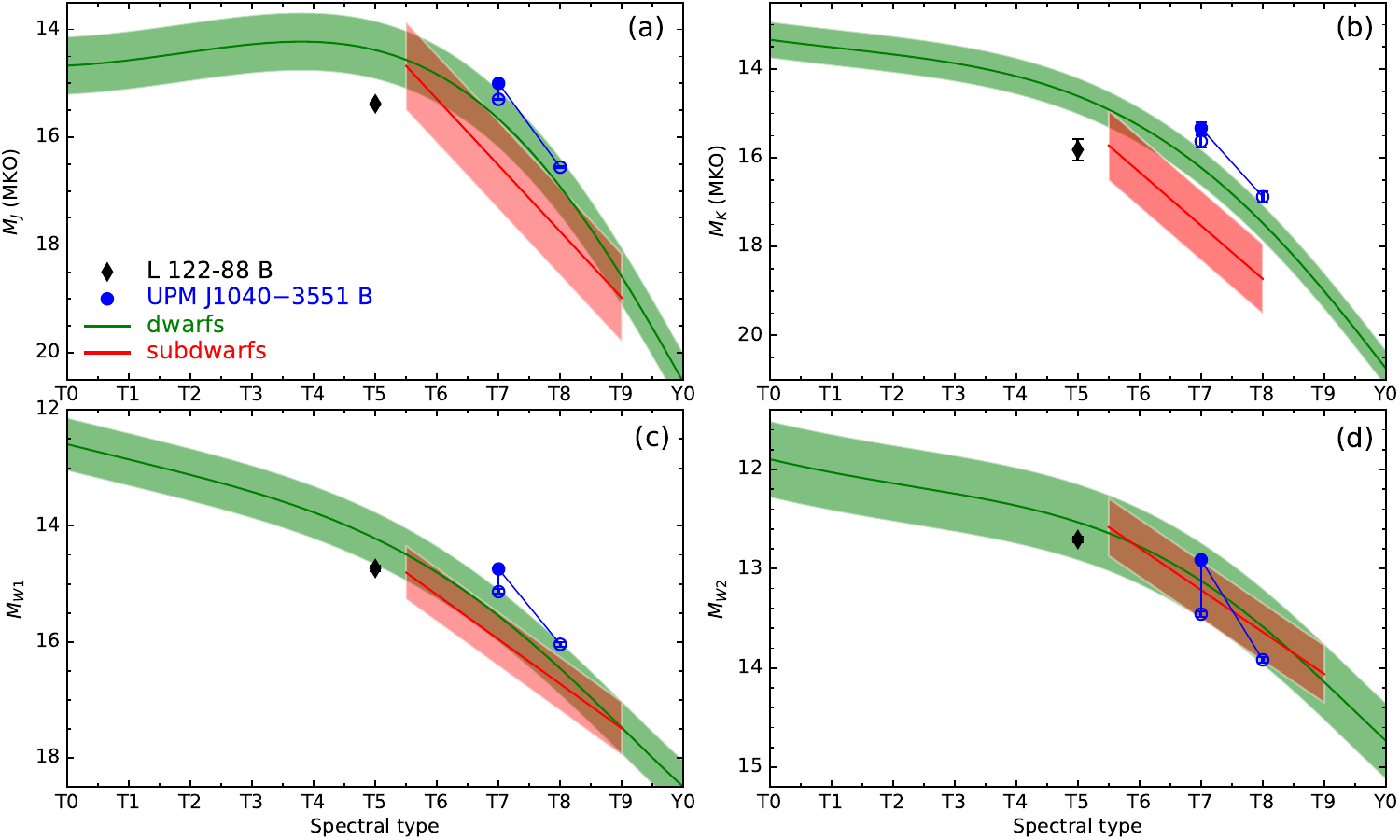}
    \caption{
    Absolute magnitudes ($M_J$, $M_K$, $M_{W1}$ and $M_{W2}$) of L 122-88 B (black diamond) and UPM J1040$-$3551 B (blue circle) plotted against spectral type. Long curved green and short straight red lines show the spectral type versus absolute magnitude correlations for dwarfs and subdwarfs, respectively, from fig. 10 of \citet{prime6}. The green shaded areas indicate the rms of these fits. Blue open circles represent the decomposed absolute magnitudes of UPM J1040$-$3551 Ba \& Bb, obtained by splitting the magnitudes of UPM J1040$-$3551 B based on the average differences between T7 and T8 dwarfs.
    }
    \label{fspt}
\end{figure*}

\begin{figure}
    \includegraphics[width=\columnwidth]{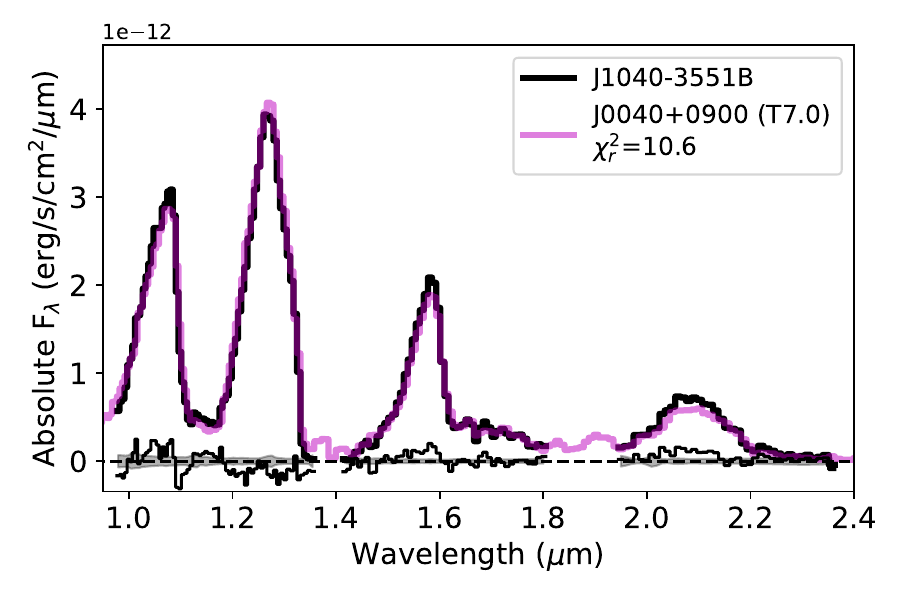} \\
    \includegraphics[width=\columnwidth]{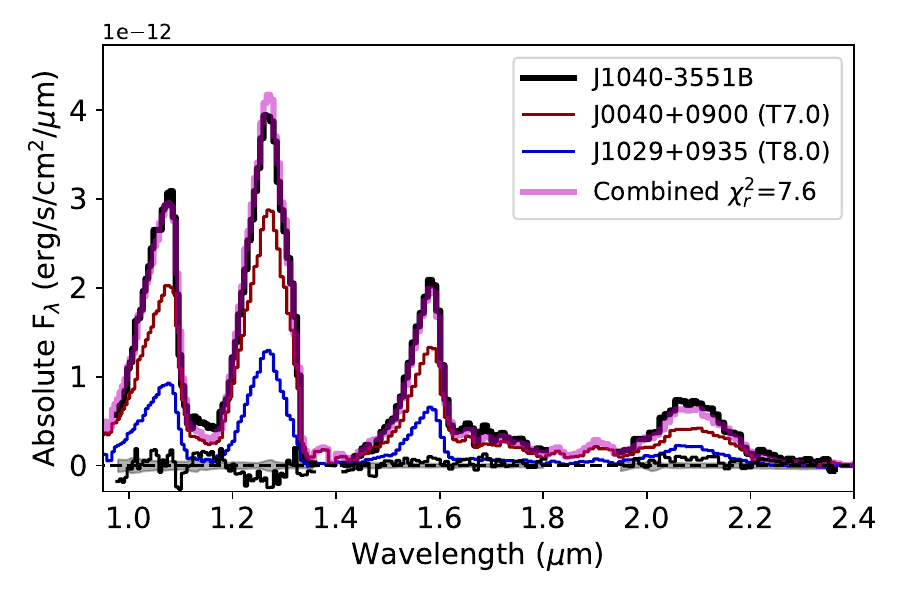}   
    \caption{SOAR spectrum of UPM J1040$-$3551 B (black lines) scaled to absolute fluxes compared to best-fitting single (top panel) and binary templates (bottom panel; magenta lines) constructed from SpeX/prism spectra. 
    In the bottom panel, we display also the relative scaling of the spectral components of the best-fitting binary, the T7 J0040+0900 (red line) and T8 J1029+0935 (blue line). Both panels compare the difference spectrum (black line centred at zero) to the uncertainty spectrum (grey band), and the reduced $\chi^2$ is listed for each fit.}
    \label{fbfit}
\end{figure}

\begin{figure}
	\includegraphics[width=\columnwidth]{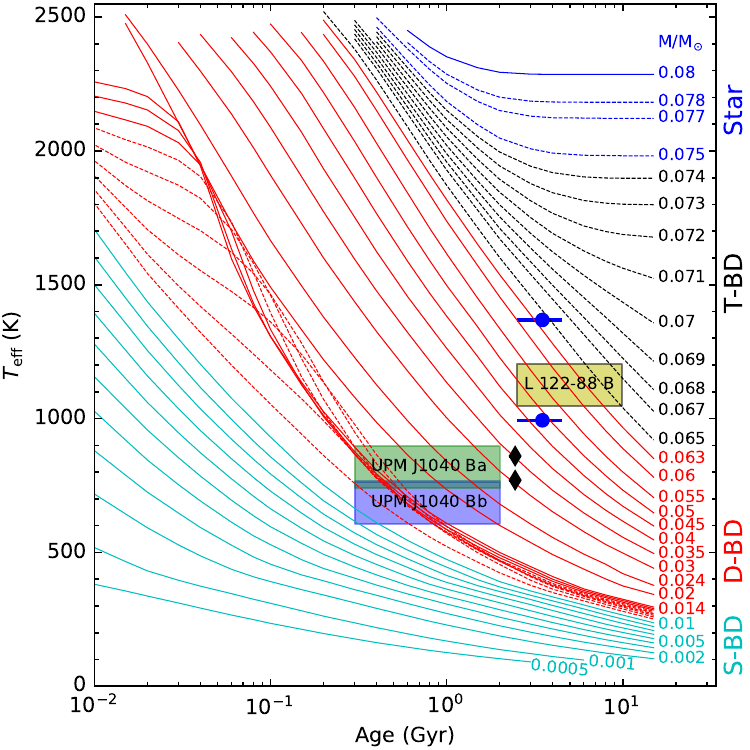}
    \caption{Evolutionary tracks of $T_{\rm eff}$ from the Sonora-Bobcat models \citep{marl21} for objects with masses of 0.0005--0.08 M$_{\odot}$ and solar metallicity, spanning 0.02--15 Gyr. Tracks are colour-coded from top to bottom: blue for stars, black for transitional BDs \citep[T-BDs,][]{prime3}, red for degenerate BDs \citep[D-BDs,][]{prime6}, and cyan for sub-brown dwarfs (S-BDs). Yellow, green and blue shaded boxes indicate the estimated locations of L 122-88 B and UPM J1040$-$3551 Ba \& Bb, respectively. For comparison, $\varepsilon$ Indi BaBb (blue circles) and Gliese 229 BaBb (black diamonds) are also plotted. Note that stars with masses of 0.075--0.078 M$_{\odot}$ (blue dashed lines) keep reducing their $T_{\rm eff}$ and radii slowly until 8-10 Gyr according to the Sonora-Bobcat models. D-BDs with masses between 0.01 and 0.014 M$_{\odot}$ (red dashed lines) fuse partial of their deuterium and form a lithium burning transition zone \citep[see fig. 5 of][]{prime6}. }
    \label{fevo}
\end{figure}

\section{Characterization}
\label{scha}
\subsection{M dwarf companions}
\label{smdc}
The {\sl Gaia} Renormalized Unit Weight Error (RUWE) values for L 122-88 A and UPM J1040$-$3551 A (1.32 and 1.47, respectively) suggest that these objects may potentially be unresolved binaries. However, their optical spectra appear consistent with normal M dwarfs (Fig. \ref{fspeca}), indicating that if they are indeed binaries, each component likely possesses a similar spectral classification. Under such circumstances, given their identical distances, metallicity constraints derived from model fitting, whether applied to their composite spectra or to individual components, would yield consistent results. In this section, we first examine their spectra as single objects, then discuss their binarity in Section \ref{ssemb}.

\subsubsection{L 122-88 A}
\label{sj0031a}
We compared the optical spectra (Fig. \ref{fspeca}) of L 122-88 A to those M0-M9 dwarf templates from \citet{boch07}, and found that it closely matches an M2 dwarf template. However, it exhibits slightly weaker TiO absorption bands around 710 nm, which indicates a mildly metal-poor composition \citep[e.g.,][]{lepi07,jao08,prime7}.

Analysis of data from the {\sl Transiting Exoplanet Survey Satellite} \citep[{\sl TESS};][]{tess} using the {\scriptsize TESSILATOR} software \citep{binks24} reveals a rotation signal of 2.475 $\pm$ 0.013 d in sector 2 data. Applying gyrochronology to this rotation period using GPgyro \citep{lu24} yields an age estimate of 0.029$\pm$0.010 Gyr. However, there is no evidence for this signal in other {\sl TESS} sectors (28, 29, 68, and 69). Examination of the {\sl TESS} data sector 2 release notes\footnote{\url{https://ntrs.nasa.gov/citations/20190001284}} for TIC 271503151 (L 122-88 A) does not suggest any particular issue with the sector 2 observation but for a $V$ = 13 object we would expect the signal to be visible in other sectors for a reliable detection. \citet{maro24} suggested a significantly older kinematic age of 2.5-9.9 Gyr for L 122-88 A.

\subsubsection{UPM J1040$-$3551 A}
\label{sj1040a}

We compared the optical spectrum (Fig. \ref{fspeca}) of UPM J1040$-$3551 A to those M0--M9 dwarf templates from \citet{boch07}, and found it closely matches that of an M4 dwarf template. The presence of H$\alpha$ emission in its spectrum indicates that it is an active M4 dwarf, which is also suggestive of a relatively young age. We measured the H$\alpha$ equivalent width (H$\alpha$ EW) of UPM J1040$-$3551 A to be between 2.33 and 3.78 \AA. Applying the H$\alpha$ activity--age correlation for M3--M6 dwarfs from \citet{kima21} (their fig. 6), this H$\alpha$ EW range corresponds to an estimated age of 0.3-2.0 Gyr. The {\sl TESS} data for UPM J1040$-$3551 A show no indication of activity or any periodic signal.

\subsubsection{Atmospheric properties}
\label{satmos}

The optical spectrum is more sensitive to $T_{\rm eff}$ of M (sub)dwarfs, while the NIR spectrum is more sensitive to [Fe/H]. Therefore, the full optical--NIR spectral profile of M subdwarfs can have better constraints of their $T_{\rm eff}$ and [Fe/H]. To derive the atmospheric parameters of L 122-88 A, we fitted its X-shooter spectrum (300--2500 nm) to BT-Settl models with 2600 K$ \leq T_{\rm eff} \leq 4500$ K, $-1.0 \leq$ [Fe/H] $\leq$ 0.5, and log $g$ = 5.0 \citep{alla14}. We used a typical log $g$ = 5.0 for M2--M4 dwarfs as log $g$ has relatively small effect on their spectra compared to $T_{\rm eff}$ and [Fe/H]. We applied smoothing to the X-shooter spectrum to enhance the SNR, though this degraded the resolving power to $\sim$1000. The BT-Settl model spectra were similarly degraded to match the resolution of the processed X-shooter spectrum.

The fitting procedure involved comparing the observed spectrum with a grid of models, varying $T_{\rm eff}$ in steps of 50 K and [Fe/H] in steps of 0.1 dex. We employed a least-squares method for the comparison. The best-fitting model for L 122-88 A yielded $T_{\rm eff}$ = 3500 $\pm$ 50 K, [Fe/H] = $-0.2 \pm 0.1$, and log $g$ = 5.0 (Fig. \ref{fj0031axs}), confirming its mildly metal-poor nature.

We applied a similar fitting procedure to the optical spectrum of UPM J1040$-$3551 A, using the same set of BT-Settl models. The best-fitting model for UPM J1040$-$3551 A resulted in $T_{\rm eff}$ = 3200 K, [Fe/H] = 0.0, and log $g$ = 5.0 (Fig. \ref{fspeca}), indicating a solar metallicity for this object. 

\subsubsection{UPM J1040$-$3551 A: a likely near equal-mass binary}
\label{ssemb}

Single stars are expected to have RUWE $\approx$ 1.0. Therefore, RUWE $>$ 1.4 is usually used as an indicator of non-single object. Nearby M dwarfs and earlier type stars predominantly exhibit RUWE values of 1.2 $\pm$ 0.3 and 1.0 $\pm$ 0.2, respectively \citep{sozz23}. The RUWE = 1.47 of UPM J1040$-$3551 A suggests it is likely an unresolved binary. For unresolved equal-mass binaries, the combined magnitude appears $\sim$0.75 mag brighter than each individual component, providing a method to test the binarity hypothesis.

UPM J1040$-$3551 A exhibits a typical $T_{\rm eff}$ (3200 $\pm$ 50 K) for M4 dwarfs \citep[see fig. 4 of][]{prime3}. Notably, model spectral fitting would yield identical temperature estimates regardless of whether the object is a single star or an equal-mass binary. According to evolutionary models \citep{bara15}, a solar-metallicity star with $T_{\rm eff} = 3200$ K and age of 0.5-5 Gyr should have a mass of 0.174 M$_{\odot}$, radius of 0.195 R$_{\odot}$, and absolute magnitudes of $M_J = 8.98$, $M_H = 8.39$, and $M_K = 8.13$. The observed absolute magnitudes of UPM J1040$-$3551 A ($M_J = 8.29$, $M_H = 7.66$, and $M_K = 7.432$) are systematically brighter by 0.69, 0.73, and 0.70 mag, respectively, than predicted by models for its temperature (see Fig. \ref{ftjhk}). This consistent $\sim$0.7 mag excess strongly supports the classification of UPM J1040$-$3551 A as a near equal-mass binary system (UPM J1040$-$3551 Aab), with components Aa and Ab differing by only $\sim$0.1 mag ($\Delta T_{\rm eff} \sim$ 25 K) and both can be classified as M4 dwarfs.

As a single star, UPM J1040$-$3551 A would have a mass of $0.235 \pm 0.020$ M$_{\odot}$ according to the mass-$M_K$ correlation \citep{mann19}. A solar-metallicity star with this mass and age of 0.5-5 Gyr should have a $T_{\rm eff} = 3320 \pm 40$ K, significantly higher than the model spectral fitting result. With $T_{\rm eff}$ of 3200 K and 3175 K, UPM J1040$-$3551 Aa \& Ab should have a mass around 0.174 $\pm$ 0.022 M$_{\odot}$ and 0.165 $\pm$ 0.020 M$_{\odot}$, respectively, according to evolutionary models \citep{bara15}.

For comparison, we also evaluated L 122-88 A by comparing its observed $M_J$, $M_H$, and $M_K$ values to evolutionary model predictions for a solar-metallicity star with $T_{\rm eff} = 3500$ K and age of 8 Gyr (see Fig. \ref{ftjhk}). The resulting magnitude differences of approximately 0.11 mag are insufficient to suggest binarity in this case.

\subsection{T dwarf companions}
\label{sstd}

Figure \ref{ftspec} presents the NIR spectra of L 122-88 B and UPM J1040$-$3551 B obtained with TripleSpec. These spectra have been smoothed to facilitate comparison with T dwarf standards. Our analysis of L 122-88 B primarily relies on its TripleSpec spectrum, which offers higher SNR compared to its X-shooter spectrum (Fig. \ref{fj0031b}).

\subsubsection{L 122-88 B}
\label{sj0031b}

The NIR spectrum of L 122-88 B (Figs \ref{ftspec} and \ref{fj0031b}) fits well to that of the T5 standard 2MASS J15031961+2525196 \citep[J1503+25;][]{burg03,burg04}, but exhibits suppressed $K$-band flux which is a subsolar metallicity feature of BDs \citep{burg02,prime1}. It bears a strong resemblance to the sdT5.5 subdwarf benchmark HIP 73786 B \citep{murr11,prime6}, which has a metallicity of $-0.30 \lesssim$ [Fe/H] $\lesssim -0.38$ inferred from its primary HIP 73786 A \citep{caro07,soub16,aren19}. The slightly weaker $K$-band flux suppression in L 122-88 B compared to HIP 73786 B suggests a marginally higher metallicity, consistent with the metallicity derived from model fitting of L 122-88 A ([Fe/H] = $0.2\pm0.1$, Section \ref{smdc}). Consequently, we classify L 122-88 B as a mildly metal-poor T5 dwarf (d/sdT5).

HIP 73786 B exhibits a weak flux excess on the left-hand side of the $Y$-band peak ($\sim$980-1080 nm), which is possibly caused by subsolar metallicity. This feature is also visible in spectra of other T subdwarfs with [Fe/H] $\lesssim -0.3$ and becomes more significant at lower metallicity \citep[e.g.,][]{pinf12,burn14}. The absence of the $Y$-band flux excess feature in L 122-88 B indicates a metallicity higher than that of T subdwarfs, which usually have [Fe/H] $\gtrsim -0.3$ \citep[e.g., fig. 9 of][]{prime2}. This is still consistent with the metallicity derived from L 122-88 A.

To further investigate the mildly metal-poor nature of L 122-88 B, we compared its absolute magnitudes in the $J, K, W1,$ and $W2$ bands, computed using the {\sl Gaia} parallax of the primary, to those of known T dwarfs and subdwarfs (Fig. \ref{fspt}). L 122-88 B exhibits slightly fainter $M_J, M_K,$ and $M_{W1}$ compared to typical T5 dwarfs, while its $M_{W2}$ remains similar. This pattern is consistent with the behaviour of T5.5-9 subdwarfs, which show fainter $Y, J, H, K,$ and $W1$-band absolute magnitudes than their solar-metallicity counterparts, but indistinguishable $W2$-band absolute magnitudes \citep{prime6}. Without considering its mildly metal-poor nature, L 122-88 B might be misclassified as a T6.5 $\pm$ 1 dwarf based on spectral type versus absolute magnitude correlations (Fig. \ref{fspt}). Indeed, \citet{maro24} classified it as a T6 dwarf based on spectral type versus colour correlations from \citet{kirk21}. 

We also estimated the distance of L 122-88 B by the spectral type -- absolute magnitude correlations \citep[fig. 10 of][]{prime6}. The average distance estimated from its $J, K, W1, W2$ magnitudes is 45$^{+21}_{-17}$ pc, which is consistent to the {\sl Gaia} distance of L 122-88 A (33.106 $\pm$ 0.014) within the uncertainty. 
Of course, L 122-88 B exhibits an over estimated spectral distance due to its mildly metal-poor composition. This metallicity deficiency causes its actual absolute magnitudes to be fainter than the average values for T5 dwarfs used in the calculation.

\subsubsection{UPM J1040$-$3551 B: a probable unresolved binary}
\label{sj1040b}
The NIR spectrum of UPM J1040$-$3551 B closely resembles that of the T7 dwarf standard 2MASS J0727182+171001 \citep[J0727+17;][]{burg02,burg06}. However, Fig. \ref{ftspec} reveals flux excesses in the $Y, H,$ and $K$ bands relative to the T7 standard when normalized at the $J$-band peak. When rescaling the spectrum of UPM J1040$-$3551 B to better align with the T7 standard across $Y$, $H$, and $K$ bands, we observe a 13\% flux suppression in the $J$ band, a phenomenon neither previously observed nor explained in any single BD. 

Several features of evidence suggest that UPM J1040$-$3551 B is not a T subdwarf: (1) it exhibits $K$-band flux excess rather than the suppression typical of late-type T subdwarfs; (2) it has a relatively young age of 0.3-2.0 Gyr inferred from its primary (Section \ref{sj1040a}); and (3) it has solar metallicity as indicated by its primary's spectrum (Fig. \ref{fspeca}). While the $K$-band flux excess could be attributed to its young age \citep[e.g.,][]{burn11}, this explanation fails to account for the observed $Y$- and $H$-band flux excesses.

Unresolved binary BDs may show unusual spectral features and colours, and flux excesses if they have been treated as single objects \citep{burg10,zhan10,maro15,prime4,kirk21}.  
Figure \ref{fspt} demonstrates that UPM J1040$-$3551 B has significantly brighter $J$-, $K$-, and $W1$-band absolute magnitudes than typical T7 dwarfs. Such overluminosity in field ultracool dwarfs often indicates the presence of an unresolved companion. 

To test this hypothesis, we conducted a binary spectral fitting analysis using single spectral templates in the Spex Prism Library Analysis Toolkit \citep[SPLAT;][]{2017ASInC..14....7B}.  We selected 131 T4--T9 dwarfs with low-resolution ($\lambda/\Delta\lambda \approx$ 150) NIR spectra previously obtained with the Spectral cross-disperser (SpeX) spectrograph \citep{2003PASP..115..362R} on the 3m NASA Infrared Telescope Facility. These templates exclude known or suspected binaries and spectra with very low S/N $\lesssim$ 15. Spectra were interpolated and resampled on to a common wavelength scale spanning 0.95--2.5~$\mu$m at a constant resolution $\lambda/\Delta\lambda$ = 200, classified by comparison to NIR spectral standards \citep{burg06}, and then flux-calibrated using the \citet{kirk21} absolute MKO $J$-band magnitude/spectral type relation. We then co-added pairs of spectra to create 6,715 unique binary templates. We compared both single and binary templates to the equivalently interpolated SOAR spectrum of J1040$-$3551 B following procedures described in \citet{burg10}, masking out regions of strong telluric absorption between 1.35--1.42~$\mu$m and 1.80--1.95~$\mu$m.
Figure~\ref{fbfit} compares the best fitting single template from this sample, the T7 dwarf WISE~J004024.88+090054.8 \citep[J0040+0900;]{mace13} to the best-fitting binary template composed of J0040+0900 and the T8 dwarf ULAS J102940.52+093514.6 \citep[J1029+0935;][]{burn13,2013PASP..125..809T}. The latter is a superior fit, notably providing a better reproduction of the $Y$, $H$, and $K$-band peaks. The difference in reduced $\chi^2$ between the single fit (10.6) and the binary fit (7.6) implies a 2\% likelihood of equivalence based on the F-test statistic. However, we caution that other physical parameters, such as youth/low surface gravity, could also explain the modest deviations in the single template fit and may not be fully accounted for in the existing SPLAT template sample. The spectral fitting therefore provides supporting but not conclusive evidence of unresolved multiplicity.

To further investigate this binary hypothesis, we decomposed the absolute magnitudes of UPM J1040$-$3551 B into T7 and T8 components. Using the spectral type versus absolute magnitude correlations for L and T dwarfs (Fig. \ref{fspt}), we derived $M_J$, $M_K$, $M_{W1},$ and $M_{W2}$ differences between T7 and T8 dwarfs of 1.258, 1.248, 0.907, and 0.462 mag, respectively.
The decomposed $M_J$, $M_K$, and $M_{W1}$ values demonstrate improved alignment with these correlations, though they remain slightly brighter than average -- contrary to the dimmer characteristics of late T subdwarfs. 
UPM J1040$-$3551 B appears underluminous as an unresolved binary only in the $W2$ band, which may be attributed to its young age. For instance, the young T5.5 dwarf SDSSp J111010.01+011613.1 \citep{geba02} is relatively faint in the $W2$ band, with a $W1-W2$ colour 0.34 mag bluer than the average for field T5.5 dwarfs \citep[fig. 8i of][]{prime6}.

The independent spectroscopic distance of UPM J1040$-$3551 B as a single T7 dwarf is 20$^{+9}_{-8}$ pc, which is 21 percent shorter than the {\sl Gaia} distance of the M4 primary (25.283$\pm$0.013 pc). We also decomposed the $J$-, $K$-, $W1$-, $W2$-band apparent magnitudes of UPM J1040$-$3551 B as T7 + T8 close binary. The average spectroscopic distance of UPM J1040$-$3551 Ba as a T7 dwarf is 21$^{+10}_{-8}$ pc, which is 17 percent smaller than its UPM J1040$-$3551 A but consistent within the uncertainty. 
The spectral distance of UPM J1040$-$3551 B is under estimated due to its young age, which allows it to have brighter absolute magnitudes than the average values for T7 dwarfs used in the calculation.   

Both the spectral fitting and the decomposition of absolute magnitudes strongly suggest that UPM J1040$-$3551 B is likely an unresolved T7 + T8 binary (UPM J1040$-$3551 Ba \& Bb). UPM J1040$-$3551 AabBab is the  first known quadruple system comprising a close binary of T dwarfs in a wide orbit around a close binary of stars. It is the third identified system featuring a close binary of T dwarfs with wide stellar companion, following $\varepsilon$ Indi ABab \citep{scho03,mcca04,king10} and Gliese 229 ABab \citep{xuan24}.

UPM J1040$-$3551 B is detected but not resolved in the Visible and Infrared Survey Telescope for Astronomy (VISTA) Hemisphere Survey \citep[VHS;][]{mcma13} and the Dark Energy Camera Legacy Survey \citep[DECaLS;][]{dey19}. The VHS and DECaLS images have pixel sizes of 0.34 and 0.26 arcsec, respectively. This allows us to place upper limits on the orbital semimajor axis ($\lesssim 10$ au) and period ($\lesssim 160$ yr) of the system. At a distance of 25 pc, the components can potentially be resolved at 1.15 $\mu$m by the Near-Infrared Camera of the {\sl JWST}, provided the orbital period exceeds six years.

\subsection{Binary properties}
\label{ssbi}

To estimate the masses of the T dwarf companions, we employed the evolutionary tracks of $T_{\rm eff}$ from the Sonora-Bobcat models \citep{marl21}, using their estimated ages and the average $T_{\rm eff}$ of T dwarfs with corresponding spectral types. As illustrated in Fig. \ref{fevo}, L 122-88 B has an estimated mass of 0.05--0.067 M${\odot}$ for an age of 2.5-9.9 Gyr, while UPM J1040$-$3551 Ba and Bb have estimated masses of 0.012--0.04 M${\odot}$ and 0.007--0.02 M$_{\odot}$, respectively, for an age of 0.3-2.0 Gyr.
Using the derived distances, projected separations, and masses of all components, we calculated the binding energies of L 122-88 AB and UPM J1040$-$3551 AabBab to be $5.6\times10^{33}$ and $1.3\times10^{34}$ J, respectively.

To assess the stability of these wide binary systems, we considered the Jacobi (or tidal) radius ($r_{\rm J}$), which represents the maximum separation for stable wide binary systems. The $r_{\rm J}$ delineates the boundary beyond which the tidal field dominates over the gravitational attraction between the binary components. For wide binaries in the solar neighbourhood, $r_{\rm J}$ can be calculated using equation (43) from \citet{jian10}.

We found that L 122-88 AB and UPM J1040$-$3551 AabBab have $r_{\rm J}$ values of $\sim$1.02 and $\sim$0.97 pc, respectively. Their projected separations correspond to 0.034 $r_{\rm J}$ and 0.008 $r_{\rm J}$, respectively, which are far smaller than their respective $r_{\rm J}$ values. This analysis strongly suggests that both systems are gravitationally stable. The spectroscopic distances of L 122-88 B and UPM J1040$-$3551 B are also consistent to {\sl Gaia} parallax distances of their primaries (see Table \ref{ttb}).

\section{Summary and conclusions}
\label{scon}
We have conducted a search for late T dwarf wide binaries using CatWISE and {\sl Gaia} DR3, resulting in the discovery of a wide binary and a probable hierarchical quadruple system. Both systems are confirmed to be gravitationally bound and stable based on their tidal radii. Optical and NIR spectroscopic follow-up observations were performed for all companions, enabling spectral classification and characterization.

The first system, L 122-88 AB, exhibits spectral features indicative of mildly metal-poor composition, including a weaker 710 nm TiO absorption band in L 122-88 A and suppressed $K$-band flux in L 122-88 B. The best-fitting BT-Settl model to the 0.3-2.5 $\mu$m spectrum of L 122-88 A yields $T_{\rm eff}$ = 3500 K, [Fe/H] = $-0.2$, and log $g$ = 5.0. We classify L 122-88 AB as a mildly metal-poor M2 + T5 wide binary.

The second system, UPM J1040$-$3551 AB, is likely a hierarchical quadruple system consisting of a T7 + T8 spectral binary and an  M4 + M4 astrometric binary at a distance of 25 pc from the Sun. Both the RUWE (1.47) of UPM J1040$-$3551 A and its excess luminosity (0.7 mag brighter than model predications of its $T_{\rm eff}$) suggest it is an unresolved M4 + M4 binary. While UPM J1040$-$3551 Ba/Bb are unresolved in survey images, their NIR spectrum is best fitted by a combined T7 + T8 template, explaining the observed excess luminosity. UPM J1040$-$3551 AabBab is the first known quadruple system with a close binary of T dwarfs. Based on the H$\alpha$ emission of the M4 primary, we estimate the system age to be 0.3-2 Gyr. At this relatively young age, the T8 component UPM J1040$-$3551 Bb has a mass between 9 and 28 Jupiter masses and potentially is a planetary mass object. 

These newly discovered systems, L 122-88 AB and UPM J1040$-$3551 AB, augment the sample of benchmark BDs, particularly in the less-explored regimes of cool temperature. UPM J1040$-$3551 B potentially is a good target for dynamical mass measurement with high resolution imaging. They present valuable opportunities for refining atmospheric and evolutionary models, thereby enhancing our understanding of substellar objects approaching the planetary mass limit. Furthermore, these systems may provide crucial insights into bridging the gap between BDs and giant exoplanets, contributing to a more comprehensive understanding of the low-mass end of the stellar-substellar-planetary continuum.

\section*{Acknowledgements}
This study is based on observations obtained at the SOAR telescope, which is a joint project of the Minist\'{e}rio da Ci\^{e}ncia, Tecnologia e Inova\c{c}\~{o}es (MCTI/LNA) do Brasil, the US National Science Foundation’s NOIRLab, the University of North Carolina at Chapel Hill (UNC), and Michigan State University (MSU). 
This study is based on observations collected at the European Southern Observatory under ESO programme 113.26V7.001.

This work has made use of data from the European Space Agency (ESA) mission {\it Gaia} (\url{https://www.cosmos.esa.int/gaia}), processed by the {\it Gaia} Data Processing and Analysis Consortium (DPAC, \url{https://www.cosmos.esa.int/web/gaia/dpac/consortium}). Funding for the DPAC has been provided by national institutions, in particular the institutions participating in the {\sl Gaia} Multilateral Agreement.
This publication makes use of data products from the {\sl WISE}, which is a joint project of the University of California, Los Angeles, and the Jet Propulsion Laboratory/California Institute of Technology, funded by the National Aeronautics and Space Administration. 
SPLAT is a collaborative project of research students in the UCSD Cool Star Lab, aimed at developing research through the building of spectral analysis tools. Contributors to SPLAT have included Christian Aganze, Jessica Birky, Daniella Bardalez Gagliuffi, Adam Burgasser (PI), Caleb Choban, Andrew Davis, Ivanna Escala, Joshua Hazlett, Carolina Herrara Hernandez, Elizabeth Moreno Hilario, Aishwarya Iyer, Yuhui Jin, Mike Lopez, Dorsa Majidi, Diego Octavio Talavera Maya, Alex Mendez, Gretel Mercado, Niana Mohammed, Johnny Parra, Maitrayee Sahi, Adrian Suarez, Melisa Tallis, Tomoki Tamiya, Chris Theissen, and Russell van Linge.

ZHZ acknowledges the supports from the Jiangsu Province fundamental research programme (BK20211143), the Program for Innovative Talents and Entrepreneur in Jiangsu (JSSCTD202139), and the science research grants from the China Manned Space Project with no. CMS-CSST-2021-A08.
The work of FN is supported by NOIRLab, which is managed by the Association of Universities for Research in Astronomy (AURA) under a cooperative agreement with the National Science Foundation.
MCGO acknowledges financial support from the Agencia Estatal de Investigación (AEI/10.13039/501100011033) of the Ministerio de Ciencia e Innovación and the ERDF 'A way of making Europe' through project PID2022-137241NB-C42. 
NL acknowledges support from the Agencia Estatal de Investigaci\'on del Ministerio de Ciencia e Innovaci\'on (AEI-MCINN) under grant PID2022-137241NB-C41. BG acknowledges support from the Polish National Science Center (NCN) under SONATA grant no. 2021/43/D/ST9/0194. PC acknowledges financial support from the Spanish Virtual Observatory project funded by the Spanish Ministry of Science and Innovation/State Agency of Research MCIN/AEI/10.13039/501100011033 through grant PID2020-112949GB-I00. The authors would like to thank the anonymous reviewer for his/her valuable comments.

%The Acknowledgements section is not numbered. Here you can thank helpful colleagues, acknowledge funding agencies, telescopes and facilities used etc.
%Try to keep it short.

%%%%%%%%%%%%%%%%%%%%%%%%%%%%%%%%%%%%%%%%%%%%%%%%%%
\section*{Data Availability}
The data underlying this article are available in the article and in its online supplementary material.

%Check Sample Data Availability Statements at \url{https://academic.oup.com/pages/open-research/research-data#Data%20Availability%20Statements}.

%The inclusion of a Data Availability Statement is a requirement for articles published in MNRAS. Data Availability Statements provide a standardised format for readers to understand the availability of data underlying the research results described in the article. The statement may refer to original data generated in the course of the study or to third-party data analysed in the article. The statement should describe and provide means of access, where possible, by linking to the data or providing the required accession numbers for the relevant databases or DOIs.

%%%%%%%%%%%%%%%%%%%% REFERENCES %%%%%%%%%%%%%%%%%%

% The best way to enter references is to use BibTeX:

\bibliographystyle{mnras}
\bibliography{example} % if your bibtex file is called example.bib

% Alternatively you could enter them by hand, like this:
% This method is tedious and prone to error if you have lots of references
%\begin{thebibliography}{99}
%\bibitem[\protect\citeauthoryear{Author}{2012}]{Author2012}
%Author A.~N., 2013, Journal of Improbable Astronomy, 1, 1
%\bibitem[\protect\citeauthoryear{Others}{2013}]{Others2013}
%Others S., 2012, Journal of Interesting Stuff, 17, 198
%\end{thebibliography}

%%%%%%%%%%%%%%%%%%%%%%%%%%%%%%%%%%%%%%%%%%%%%%%%%%

%%%%%%%%%%%%%%%%% APPENDICES %%%%%%%%%%%%%%%%%%%%%

%\appendix

%\section{Some extra material}

%If you want to present additional material which would interrupt the flow of the main paper,
%it can be placed in an Appendix which appears after the list of references.

%%%%%%%%%%%%%%%%%%%%%%%%%%%%%%%%%%%%%%%%%%%%%%%%%%

% Don't change these lines
\bsp	% typesetting comment
\label{lastpage}
\end{document}